\documentclass[12pt]{article}

\title{Hamiltonian and Brownian systems with long-range interactions: III. The BBGKY hierarchy for spatially inhomogeneous systems}

\usepackage{amsthm,amsmath,amssymb,a4wide,epsf}

cm \hsize=18 true cm \topmargin=-2cm \oddsidemargin=-0.5cm
\def\mb#1{\setbox0=\hbox{$#1$}\kern-.025em\copy0\kern-\wd0
\kern-0.05em\copy0\kern-\wd0\kern-.025em\raise.0233em\box0}

\begin{document}

\author{Pierre-Henri Chavanis}
\maketitle
\begin{center}
Laboratoire de Physique Th\'eorique (CNRS UMR 5152), \\
Universit\'e
Paul Sabatier,\\ 118, route de Narbonne, 31062 Toulouse Cedex 4, France\\
E-mail: {\it chavanis{@}irsamc.ups-tlse.fr\\
 }
\vspace{0.5cm}
\end{center}

\begin{abstract}

We study the growth of correlations in systems with weak long-range
interactions. Starting from the BBGKY hierarchy, we determine the
evolution of the two-body correlation function by using an expansion
of the solutions of the hierarchy in powers of $1/N$ in a proper
thermodynamic limit $N\rightarrow +\infty$.  These correlations are
responsible for the ``collisional'' evolution of the system beyond the
Vlasov regime due to finite $N$ effects. We obtain a general kinetic
equation that can be applied to spatially inhomogeneous systems and
that takes into account memory effects. These peculiarities are
specific to systems with unshielded long-range interactions. For
spatially homogeneous systems with short memory time like plasmas, we
recover the classical Landau (or Lenard-Balescu) equations.  An
interest of our approach is to develop a formalism that remains in
physical space (instead of Fourier space) and that can deal with
spatially inhomogeneous systems. This enlightens the basic physics and
provides novel kinetic equations with a clear physical
interpretation. However, unless we restrict ourselves to spatially
homogeneous systems, closed kinetic equations can be obtained only if
we ignore some collective effects between particles. General exact
coupled equations taking into account collective effects are also
given. We use this kinetic theory to discuss the processes of violent
collisionless relaxation and slow collisional relaxation in systems
with weak long-range interactions.  In particular, we investigate the
dependence of the relaxation time with the system size and
provide a coherent discussion of all the numerical results obtained
for these systems.

\end{abstract}

\section{Introduction}
\label{sec_introduction}

Systems with long-range interactions are numerous in nature
\cite{dauxois}. Some examples include self-gravitating systems,
two-dimensional vortices, neutral and non-neutral plasmas, bacterial
populations, defects in solids, etc... When the potential of
interaction is attractive and unshielded, these systems can
spontaneously organize into {\it coherent structures} accounting for
the diversity of the objects observed in the universe. For example,
self-gravitating systems organize into planets, stars, galaxies,
clusters of galaxies... On the other hand, two-dimensional turbulent
flows organize into jets (like the gulf stream on the earth) or
large-scale vortices (like Jupiter's great red spot in the jovian
atmosphere).  Biological populations (like bacteria, amoebae,
endothelial cells,...) also interact via long-range signals through
the phenomenon of chemotaxis. Chemotactic aggregation leads to the
spontaneous appearance of patterns like stripes and spots, filaments,
vasculature,... Although these astrophysical, hydrodynamical and
biological systems are physically different, they share a lot of
analogies due to the long-range attractive nature of the potential of
interaction
\cite{cras}.

In view of the complexity of these systems, it is natural to try to
understand their structure and organization in terms of statistical
mechanics \cite{dauxois}. Since systems with long-range interactions
are generically spatially inhomogeneous, it is clear at first sights
that the usual thermodynamic limit $N\rightarrow +\infty$ with $N/V$
fixed is not valid. Therefore, the ordinary methods of statistical
mechanics and kinetic theory must be reformulated and adapted to these
systems. We shall assume, however, that the basic concepts are not
altered so that the description of these systems must be done in
consistency with the foundations of statistical mechanics and kinetic
theory. In previous papers of this series \cite{hb1,hb2} (denoted Papers I and
II), we have undertaken a systematic study of the dynamics and
thermodynamics of systems with long-range interactions. In Paper I, we
have considered the statistical equilibrium states and the static
correlation functions. We have shown that there exists a critical
temperature $T_c$ (for Brownian systems) or a critical energy $E_c$
(for Hamiltonian systems) above which the system is spatially
homogeneous and below which the homogeneous phase becomes unstable and
is replaced by a clustered phase. In Paper II, using an analogy with
plasma physics, we have developed a kinetic theory of systems with
long-range interactions in the homogeneous phase.  In the present
paper (Paper III), we propose new derivations of the kinetic equations
that take into account non-markovian effects and that can be applied
to spatially {\it inhomogeneous} configurations. These extensions are
specific to systems with unshielded long-range interactions and they
are novel with respect to the much more studied case of neutral
plasmas. They complete the results of Paper II that were only valid
for spatially homogeneous and markovian systems. However, a limitation
of the present approach is to neglect collective effects. These
effects were taken into account in Paper II for spatially homogeneous
systems.

This paper is organized as follows.  In Sec. \ref{sec_bbgky}, we
derive a general kinetic equation for Hamiltonian systems with weak
long-range interactions from the BBGKY hierarchy. This equation is
valid at order $O(1/N)$ in an expansion of the solutions of the
equations of the hierarchy in powers of $1/N$ in the proper
thermodynamic limit $N\rightarrow +\infty$ defined in Paper I. For
$N\rightarrow +\infty$, this kinetic equation reduces to the Vlasov
equation. At order $O(1/N)$ it takes into account the effect of
``collisions'' (more properly ``correlations'') between particles due
to finite $N$ effects (graininess). It describes therefore the
evolution of the system on a timescale $Nt_D$, where $t_D$ is the
dynamical time. This general kinetic equation applies to systems that
can be spatially inhomogeneous and takes into account non-markovian
effects. If we restrict ourselves to spatially homogeneous systems and
neglect memory terms, we recover the Landau equation as a special
case. In Secs. \ref{sec_qss} and \ref{sec_dis}, we use this kinetic
theory to discuss the processes of violent collisionless relaxation
and slow collisional relaxation in systems with weak long-range
interactions. We review several results obtained for self-gravitating
systems, two-dimensional vortices and the HMF model, emphasize their
connections and try to explain them in the light of the kinetic
theory.  In particular, we investigate the dependence of the
relaxation time with the system size. We also propose a scenario
according to which, for a large class of initial conditions, the
transient states of the collisional relaxation of the HMF model could
be described by spatially homogeneous Tsallis distributions
(polytropes) with a compact support and with an index $q(t)\ge 1$
slowly decreasing with time until they become Vlasov unstable and
relax towards the Boltzmann distribution.

\section{Kinetic equation from the BBGKY hierarchy}
\label{sec_bbgky}

In this section, we derive a general kinetic equation (\ref{g4}) for Hamiltonian systems with weak long-range interactions. We start from
the BBGKY hierarchy and use a systematic expansion of the solutions
of the equations of this hierarchy in powers of $1/N$ in a proper
thermodynamic limit $N\rightarrow +\infty$. The kinetic equation (\ref{g4})
is valid at order $O(1/N)$.

\subsection{The $1/N$ expansion}
\label{sec_e}

We consider a system of $N$ particles with long-range interactions
described by the Hamiltonian equations (I-1).  Basically, the
evolution of the $N$-body distribution function is governed by the
Liouville equation (I-2). Introducing the reduced probability
distributions (I-6), we can construct the complete BBGKY hierarchy
(II-1). The first two equations of this hierarchy, governing the
evolution of the one and two-body distributions $P_{1}({\bf x}_{1},t)$
and $P_{2}({\bf x}_{1},{\bf x}_{2},t)$, are given by Eqs.  (II-2) and
(II-3). We recall that ${\bf x}$ stands for (${\bf r},{\bf v})$. We
now decompose the distributions functions in the form (I-14) and
(I-15) where $P_{2}'({\bf x}_{1},{\bf x}_{2},t)$ and $P_{3}'({\bf
x}_{1},{\bf x}_{2},{\bf x}_{3},t)$ are the two and three-body
correlation functions (or cumulants). Substituting this decomposition
in Eq. (II-2), we first obtain
\begin{eqnarray}
{\partial P_{1}\over\partial t}+{\bf v}_{1}{\partial P_{1}\over\partial {\bf r}_{1}}+(N-1){\partial P_{1}\over \partial {\bf v}_{1}}\int {\bf F}(2\rightarrow 1)P_{1}({\bf x}_{2})d{\bf x}_{2}\nonumber\\
+(N-1){\partial \over\partial {\bf v}_{1}}\int {\bf F}(2\rightarrow 1)P'_{2}({\bf x}_{1},{\bf x}_{2})d{\bf x}_{2}=0,
\label{e1}
\end{eqnarray}
 where ${\bf F}(j\rightarrow i)$ is the force
by unit of mass created by particle $j$ on particle $i$. It is
related to the potential of interaction $u_{ij}=u(|{\bf r}_i-{\bf
r}_j|)$ by
\begin{eqnarray}
{\bf F}(j\rightarrow i)=-m\frac{\partial u_{ij}}{\partial {\bf
r}_i}. \label{e2}
\end{eqnarray}
Then, substituting the  decompositions (I-14) and (I-15) in (II-3)
and using Eq. (\ref{e1}) to simplify some terms, we get 
\footnote{In Paper II, some terms were missing in Eq. (II-5) of the 
BBGKY hierarchy because we systematically took $N-1\simeq N$ and
$N-2\simeq N$ which is not correct if we consider terms of order
$O(1/N)$. }
\begin{eqnarray}
\label{e3} {\partial P_{2}'\over\partial t}+{\bf v}_{1}{\partial P_{2}'\over\partial {\bf r}_{1}}+{\bf F}(2\rightarrow 1){\partial P_{2}'\over\partial {\bf v}_{1}}+{\bf F}(2\rightarrow 1)P_{1}({\bf x}_{2}){\partial P_{1}\over\partial {\bf v}_{1}}({\bf x}_{1})\nonumber\\
-P_{1}({\bf x}_{2})\frac{\partial}{\partial{\bf v}_{1}}\int {\bf F}(3\rightarrow 1)P_{1}({\bf x}_{1})P_{1}({\bf x}_{3}) d{\bf x}_{3}
\nonumber\\
-\frac{\partial}{\partial{\bf v}_{1}}\int {\bf F}(3\rightarrow 1)P_{2}'({\bf x}_{1},{\bf x}_{3})P_{1}({\bf x}_{2}) d{\bf x}_{3}\nonumber\\
+(N-2){\partial\over\partial {\bf v}_{1}}\int {\bf F}(3\rightarrow 1)P_{2}'({\bf x}_{1},{\bf x}_{2})P_{1}({\bf x}_{3})d{\bf x}_{3}\nonumber\\
+(N-2){\partial\over\partial {\bf v}_{1}}\int {\bf F}(3\rightarrow 1)P_{2}'({\bf x}_{2},{\bf x}_{3})P_{1}({\bf x}_{1})d{\bf x}_{3}\nonumber\\
+(N-2){\partial\over\partial {\bf v}_{1}}\int {\bf F}(3\rightarrow 1)P_{3}'({\bf x}_{1},{\bf x}_{2},{\bf x}_{3})d{\bf x}_{3}+(1\leftrightarrow 2)=0.
\end{eqnarray}
Equations (\ref{e1})-(\ref{e3}) are exact for all $N$ but the
hierarchy is not closed. We shall now consider the thermodynamic limit
defined in Paper I. It corresponds to $N\rightarrow +\infty$ in such a
way that the normalized temperature $\eta=\beta Nm^{2}u_{*}$ and the
normalized energy $\epsilon=E/(u_{*}N^{2}m^{2})$ are fixed, where
$u_{*}$ represents the typical strength of the potential of
interaction. In general, the potential of interaction is written as
$u({\bf r}_{ij})=k\tilde{u}({\bf r}_{ij})$ where $k$ is the coupling
constant (e.g., $G$ for self-gravitating systems or $k$ for the HMF
model). By a suitable normalization of the parameters, this
thermodynamic limit is such that the coupling constant behaves like
$k\sim u_{*} \sim 1/N$ while the individual mass $m\sim 1$, the
inverse temperature $\beta\sim 1$, the energy per particle $E/N\sim 1$
and the volume $V\sim 1$ are of order unity \footnote{Alternatively,
we can assume that the mass of the particles scales like $m\sim 1/N$
while $k\sim u_{*}\sim 1$, $\beta\sim N$, $E\sim 1$ and $V\sim 1$. In
this scaling, the total mass $M\sim Nm$ is of order unity.}. This
implies that $|{\bf x}|\sim 1$ and $|{\bf F}(j\rightarrow i)|\sim
1/N$. On the other hand, the dynamical time $t_{D}\sim R/v_{typ}\sim
1/\sqrt{k\rho}\sim 1$ is of order unity ($\rho\sim M/V$ is the average
density and the typical velocity $v_{typ}$ has been obtained by equating the
kinetic energy $\sim N m v^{2}$ and the potential energy $\sim N^2 m^{2} k
\tilde{u}$).  Since the normalized coupling constant $\beta
m^{2}u_{*}= \eta/N\sim 1/N$ goes to zero for $N\rightarrow+\infty$, we
are studying systems with {\it weak} long-range interactions. It is
argued in Papers I and II that there exists solutions of the whole
BBGKY hierarchy such that the correlation functions $P_{j}'$ scale
like $1/N^{j-1}$. This implicitly assumes that the initial condition
has no correlation, or that the initial correlations respect this
scaling (if there are strong correlations in the initial state, the
system will take a long time to erase them and the kinetic theory will
be different from the one developed in the sequel).  If this scaling
is satisfied, we can consider an expansion of the solutions of the
equations of the hierarchy in terms of the small parameter $1/N$. This
is similar to the expansion in terms of the plasma parameter made in
plasma physics. However, in plasma physics the systems are spatially
homogeneous while, in the present case, we shall take into account
spatial inhomogeneity. This brings additional terms in the kinetic
equations that are absent in plasma physics. Therefore, strictly
speaking, the hierarchy that we consider is different from the
ordinary BBGKY hierarchy. Recalling that $P_1\sim 1$, $P_2'\sim 1/N$
and $|{\bf F}(j\rightarrow i)|\sim 1/N$, we obtain at order $1/N$:
\begin{eqnarray}
{\partial P_{1}\over\partial t}+{\bf v}_{1}{\partial P_{1}\over\partial {\bf r}_{1}}+(N-1){\partial P_{1}\over \partial {\bf v}_{1}}\int {\bf F}(2\rightarrow 1)P_{1}({\bf x}_{2})d{\bf x}_{2}\nonumber\\
+N{\partial \over\partial {\bf v}_{1}}\int {\bf F}(2\rightarrow 1)P'_{2}({\bf x}_{1},{\bf x}_{2})d{\bf x}_{2}=0,
\label{e4}
\end{eqnarray}
\begin{eqnarray}
\label{e5} {\partial P_{2}'\over\partial t}+{\bf v}_{1}{\partial P_{2}'\over\partial {\bf r}_{1}}+ \left \lbrack {\bf F}(2\rightarrow 1)-\int {\bf F}(3\rightarrow 1)P_{1}({\bf x}_{3})d{\bf x}_{3}\right \rbrack P_{1}({\bf x}_{2}){\partial P_{1}\over\partial {\bf v}_{1}}({\bf x}_{1})
\nonumber\\
+N{\partial P_{2}'\over\partial {\bf v}_{1}}\int {\bf F}(3\rightarrow 1)P_{1}({\bf x}_{3})d{\bf x}_{3} +N{\partial\over\partial {\bf v}_{1}}\int {\bf F}(3\rightarrow 1)P_{2}'({\bf x}_{2},{\bf x}_{3})P_{1}({\bf x}_{1})d{\bf x}_{3}+(1\leftrightarrow 2)=0.
\end{eqnarray}
If we introduce the notations $f=NmP_{1}$ (distribution function)
and $g=N^{2}P_{2}'$ (two-body correlation function), we get
\begin{eqnarray}
{\partial f_{1}\over\partial t}+{\bf v}_{1}{\partial f\over\partial {\bf r}_{1}}+\frac{N-1}{N}\langle {\bf F}\rangle_{1} {\partial f\over \partial {\bf v}_{1}}=-m {\partial \over\partial {\bf v}_{1}}\int {\bf F}(2\rightarrow 1)g({\bf x}_{1},{\bf x}_{2})d{\bf x}_{2},
\label{e6}
\end{eqnarray}
\begin{eqnarray}
\label{e7} {\partial g\over\partial t}+{\bf v}_{1}{\partial g\over\partial {\bf r}_{1}}+\langle {\bf F}\rangle_{1} {\partial g\over\partial {\bf v}_{1}}+\frac{1}{m^{2}}
 {\bf {\cal F}}(2\rightarrow 1)f_{2}  {\partial f_{1}\over\partial {\bf v}_{1}}\nonumber\\
+\frac{\partial}{\partial {\bf v}_{1}}\int {\bf F}(3\rightarrow 1)g({\bf x}_{2},{\bf x}_{3},t) \frac{f_{1}}{m}d{\bf x}_{3}+(1\leftrightarrow 2)=0,
\end{eqnarray}
where we have introduced the abbreviations $f_{1}=f({\bf r}_{1},{\bf
v}_{1},t)$ and $f_{2}=f({\bf r}_{2},{\bf v}_{2},t)$. We have also
introduced the  mean force (by unit of mass) created in ${\bf r}_1$ by all the
particles
\begin{eqnarray}
\label{e8}
\langle {\bf F}\rangle_{1} =\int {\bf F}(2\rightarrow 1)\frac{f_2}{m}d{\bf r}_{2}d{\bf v}_{2},
\end{eqnarray}
and  the fluctuating force (by unit of mass) created by particle $2$
on particle $1$:
\begin{eqnarray}
\label{e9}
{\bf {\cal F}}(2\rightarrow 1)={\bf F}(2\rightarrow
1)-\frac{1}{N}\langle {\bf F}\rangle_{1}.
\end{eqnarray}
These equations are exact at the order $O(1/N)$. They form therefore
the right basis to develop a kinetic theory for Hamiltonian systems with
weak long-range interactions. We note that these equations are similar to
the BBGKY hierarchy of plasma physics but not identical. One
difference is the $(N-1)/N$ term in Eq. (\ref{e6}). The other
difference is the presence of the fluctuating force ${\cal
F}(2\rightarrow 1)$ instead of $F(2\rightarrow 1)$ due to the spatial
inhomogeneity of the system. In plasma physics, the system is
homogeneous over distances of the order of the Debye length so the
mean force $\langle {\bf F}\rangle$ vanishes.

\subsection{The Vlasov equation and beyond}
\label{sec_vbey}

Recalling that $P_{2}'\sim 1/N$, we note that
\begin{eqnarray}
P_{2}({\bf x}_{1},{\bf x}_{2},t)=P_{1}({\bf
x}_{1},t) P_{1}({\bf x}_{2},t)+O(1/N).
\label{e10a}
\end{eqnarray}
If we consider the limit $N\rightarrow +\infty$ (for a fixed time
$t$), we see that the correlations between particles can be neglected
so that the two-body distribution function factorizes in two one-body
distribution functions i.e. $P_{2}({\bf x}_{1},{\bf
x}_{2},t)=P_{1}({\bf x}_{1},t) P_{1}({\bf x}_{2},t)$. Therefore the
{\it mean field approximation} is exact in the limit $N\rightarrow
+\infty$. Substituting this result in Eq. (\ref{e1}), we obtain the
Vlasov equation
\begin{eqnarray}
{\partial f_{1}\over\partial t}+{\bf v}_{1}{\partial f\over\partial
{\bf r}_{1}}+\langle {\bf F}\rangle_{1} {\partial f\over \partial
{\bf v}_{1}}=0. \label{e10}
\end{eqnarray}
This equation describes the {\it collisionless evolution} of the
system up to a time at least of order $N t_D$ (where $t_D$ is the
dynamical time). In practice, $N\gg 1$ so that the domain of validity
of the Vlasov equation is huge (for example, in typical stellar systems $N\sim 10^6-10^{12}$). When the Vlasov equation is coupled to
an attractive unshielded long-range potential of interaction, it can
develop a process of violent relaxation towards a quasi stationary
state (QSS). This process will be discussed specifically in
Sec. \ref{sec_qss}.

If we want to
describe the collisional evolution of the system, we need to consider
finite $N$ effects.  Equations (\ref{e6}) and (\ref{e7}) describe the
evolution of the system on a timescale of order $N t_D$. The equation
for the evolution of the smooth distribution function is of the form
\begin{eqnarray}
{\partial f_{1}\over\partial t}+{\bf v}_{1}{\partial f\over\partial
{\bf r}_{1}}+\frac{N-1}{N}\langle {\bf F}\rangle_{1} {\partial
f\over \partial {\bf v}_{1}}=C_{N}[f], \label{e11}
\end{eqnarray}
where $C_{N}$ is a ``collision'' term analogous to the one arising in
the Boltzmann equation. In the present context, there are not real
collisions between particles. The term on the right hand side of Eq.
(\ref{e11}) is due to the development of correlations between
particles as time goes on. It is related to the two-body correlation
function $g({\bf x}_{1},{\bf x}_{2},t)$ which is itself related to the
distribution function $f({\bf x}_{1},t)$ by Eq. (\ref{e7}). Our aim is
to obtain an expression for the collision term $C_{N}[f]$ at the order
$1/N$. The difficulty with Eq. (\ref{e7}) for the two-body correlation
function is that it is an integrodifferential equation. The second
term is an advective term, the third term is the {\it source} of the
correlation and the third term takes into account the {\it
retroaction} of the system as a whole due to a change of the
correlation function. In this paper, we shall neglect the contribution
of the integral in Eq. (\ref{e7}). Then, we get the coupled system
\begin{eqnarray}
{\partial f_{1}\over\partial t}+{\bf v}_{1}{\partial f\over\partial {\bf r}_{1}}+\frac{N-1}{N}\langle {\bf F}\rangle_{1} {\partial f\over \partial {\bf v}_{1}}=-m {\partial \over\partial {\bf v}_{1}}\int {\bf F}(2\rightarrow 1)g({\bf x}_{1},{\bf x}_{2})d{\bf x}_{2},
\label{e12}
\end{eqnarray}
\begin{eqnarray}
\label{e13} {\partial g\over\partial t}+\left \lbrack {\bf v}_{1}{\partial\over\partial {\bf r}_{1}}+{\bf v}_{2}{\partial\over\partial {\bf r}_{2}} +\langle {\bf F}\rangle_{1} {\partial\over\partial {\bf v}_{1}}+\langle {\bf F}\rangle_{2} {\partial\over\partial {\bf v}_{2}}\right  \rbrack g\nonumber\\
+\left \lbrack {\bf {\cal F}}(2\rightarrow 1) {\partial\over\partial {\bf v}_{1}}+{\bf {\cal F}}(1\rightarrow 2) {\partial\over\partial {\bf v}_{2}}\right \rbrack \frac{f_{1}}{m}\frac{f_{2}}{m}=0.
\end{eqnarray}
The integral that we have neglected contains ``collective effects''
that describe the polarization of the medium. In plasma physics, they
are responsible for the Debye shielding, i.e. the fact that a charge
is surrounded by a polarization cloud of opposite charges that
diminish the interaction. These collective effects are taken into
account in the Lenard-Balescu equation through the dielectric function
(see Paper II). However, this equation is restricted to spatially
homogeneous systems and based on a Markovian approximation. These
assumptions are necessary to use Laplace-Fourier transforms in order
to solve the integro-differential equation (\ref{e7}). Here, we want
to describe more general situations where the interaction is not
shielded so that the system can be spatially inhomogeneous. If we
neglect collective effects, we can obtain a general kinetic equation
in a closed form (\ref{g4}) that is valid for systems that are not
necessarily homogeneous and that can take into account memory
effects. This equation has interest in its own right (despite its
limitations) because its structure bears a lot of physical
significance.  Before deriving this general equation, we shall first
consider the case of spatially homogeneous systems and make the link
with the familiar Landau equation.

\subsection{The Landau equation}
\label{sec_l}

For a spatially homogeneous system, the distribution function and the
two-body correlation function can be written $f=f({\bf v}_1,t)$ and
$g=g({\bf v}_1,{\bf v}_2,{\bf r}_1-{\bf r}_2,t)$. In that case,
Eqs. (\ref{e12})-(\ref{e13}) become
\begin{eqnarray}
\frac{\partial f_1}{\partial t}=m^2\frac{\partial}{\partial {\bf
v}_1}\cdot \int\frac{\partial u}{\partial {\bf x}} g({\bf v}_1,{\bf
v}_2,{\bf x},t)\, d{\bf x}d{\bf v}_2, \label{l1}
\end{eqnarray}
\begin{eqnarray}
\frac{\partial g}{\partial t}+{\bf w}\cdot \frac{\partial g}
{\partial {\bf x}}=\frac{\partial u}{\partial {\bf x}}\cdot \left
(\frac{\partial}{\partial {\bf v}_1}-\frac{\partial}{\partial {\bf
v}_2}\right )f({\bf v}_1,t)\frac{f}{m}({\bf v}_2,t),\label{l2}
\end{eqnarray}
where we have used the fact that ${\bf F}(1\rightarrow 2)=-{\bf
F}(2\rightarrow 1)$ and noted ${\bf x}={\bf r}_1-{\bf r}_2$ and ${\bf
w}={\bf v}_1-{\bf v}_2$. Taking the Fourier transform of
Eq. (\ref{l2}) and introducing the notations
$\partial={\partial}/{\partial {\bf v}_1}-{\partial}/{\partial {\bf
v}_2}$, $f_1=f({\bf v}_1,t)$ and $f_2=f({\bf v}_2,t)$, we obtain
\begin{eqnarray}
\frac{\partial \hat{g}}{\partial t}+i{\bf k}\cdot {\bf w}\hat
g=\frac{i}{m}\hat{u}(k){\bf k}\cdot\partial f_1 f_2.\label{l3}
\end{eqnarray}
In terms of the Fourier transform of the correlation function, the
kinetic equation (\ref{l1}) can be rewritten
\begin{eqnarray}
\frac{\partial f_1}{\partial t}=m^2 (2\pi)^d\frac{\partial}{\partial
{\bf v}_1}\cdot \int  {\bf k} \hat{u}(k) {\rm Im}\hat{g}({\bf
v}_1,{\bf v}_2,{\bf k},t)\, d{\bf k}d{\bf v}_2. \label{l4}
\end{eqnarray}
We shall assume that ${\rm Im}\hat{g}({\bf v}_1,{\bf v}_2,{\bf
k},t)$ relaxes on a timescale that is much smaller than the
timescale on which $f({\bf v}_1,t)$ changes. This is the equivalent
of the Bogoliubov hypothesis in plasma physics. If we ignore memory
effects, we can integrate the first order differential equation (\ref{l3})
by considering the last term as a constant. This yields
\begin{eqnarray}
\hat{g}({\bf v}_1,{\bf v}_2,{\bf k},t)=\int_0^t d\tau
\frac{i}{m}{\bf k}\hat{u}(k)e^{-i{\bf k}\cdot{\bf w}\tau}\partial
f_1(t)f_2(t),\label{l5}
\end{eqnarray}
where we have assumed that no correlation is present initially:
$g(t=0)=0$. Then, we can replace ${\rm Im}\hat{g}({\bf v}_1,{\bf
v}_2,{\bf k},t)$ in Eq. (\ref{l4}) by its value obtained for $t\rightarrow
+\infty$, which reads
\begin{eqnarray}
{\rm Im}\hat{g}({\bf k},{\bf v}_1,{\bf v}_2,+\infty)=
\frac{\pi}{m}{\bf k}\hat{u}(k)\delta({\bf k}\cdot {\bf w})\partial
f_1(t)f_2(t).\label{l6}
\end{eqnarray}
Substituting this relation in Eq. (\ref{l4}), we obtain the Landau equation
in the form
\begin{eqnarray}
\frac{\partial f_1}{\partial t}=\pi (2\pi)^d m
\frac{\partial}{\partial {v}_1^{\mu}} \int d{\bf v}_2 d{\bf k}
k^{\mu}k^{\nu} \hat{u}(k)^2 \delta({\bf k}\cdot {\bf w}) \left
(f_2\frac{\partial f_1}{\partial v_1^{\nu}}-f_1\frac{\partial
f_2}{\partial v_2^{\nu}}\right ). \label{l7}
\end{eqnarray}
Other equivalent expressions of the Landau equation are given in Paper
II. The Landau equation ignores collective effects. Collective effects
can be taken into account by keeping the contribution of the last
integral in Eq. (\ref{e7}). For spatially homogeneous systems, the
calculations can be carried out explicitly in the complex plane
\cite{ichimaru} and lead to the Lenard-Balescu equation discussed in
Paper II (the Lenard-Balescu equation can be obtained from the Landau
equation by replacing the potential $\hat{u}(k)$ by the ``screened''
potential $\hat{u}(k)/|\epsilon({\bf k},{\bf k}\cdot {\bf v}_{2})|$
including the dielectric function). The Landau and Lenard-Balescu
equations conserve mass and energy (reducing to the kinetic energy for
a spatially homogeneous system) and monotonically increase the
Boltzmann entropy
\cite{balescubook}. The collisional evolution is due to a condition of
resonance between the particles orbits. For homogeneous systems, the
condition of resonance encapsulated in the $\delta$-function appearing
in the Landau and Lenard-Balescu equations corresponds to ${\bf
k}\cdot {\bf v}_{1}={\bf k}\cdot {\bf v}_{2}$ with ${\bf v}_{1}\neq
{\bf v}_{2}$. For $d>1$, the only stationary solution is the Maxwell
distribution. Because of the $H$-theorem, the Landau and
Lenard-Balescu equations relax towards the Maxwell distribution.
Since the collision term in Eq. (\ref{l7}) is valid at order $O(1/N)$,
the relaxation time scales like
\begin{eqnarray}
t_{R}\sim Nt_D, \qquad (d>1)
 \label{l8yhy}
\end{eqnarray}
as can be seen directly from Eq. (\ref{l7}) by dimensional analysis
(comparing the l.h.s. and the r.h.s., we have $1/t_{R}\sim
u_{*}^{2}N\sim 1/N$ while $t_{D}\sim R/v_{typ}\sim 1$ with the
scalings introduced in Sec. \ref{sec_e}). A more precise estimate of
the relaxation time is given in
\cite{landaud}. For
one-dimensional systems, like the HMF model, the situation is
different. For $d=1$, the kinetic equation (\ref{l7}) reduces to
\begin{eqnarray}
\frac{\partial f_1}{\partial t}=2\pi^2 m
\frac{\partial}{\partial {v}_1} \int d{v}_2 d{k}
{k^{2}\over |k|} \hat{u}(k)^2 \delta(v_{1}-v_{2}) \left
(f_2\frac{\partial f_1}{\partial v_1}-f_1\frac{\partial
f_2}{\partial v_2}\right )=0. \label{l8}
\end{eqnarray}
Therefore, the collision term $C_{N}[f]$ vanishes at the order $1/N$
because there is no resonance. The kinetic equation reduces to
$\partial f/\partial t=0$ so that the distribution function does not
evolve at all on a timescale $\sim Nt_{D}$. This implies that, for
one-dimensional {\it homogeneous} systems, the relaxation time to statistical
equilibrium is larger than $Nt_D$. Thus, we expect that
\begin{eqnarray}
t_{R}> N t_D, \qquad (d=1).
 \label{l8yh}
\end{eqnarray}
The fact that the Lenard-Balescu
collision term vanishes in 1D is known for a long time in plasma
physics (see, e.g., the last paragraph in \cite{kp}) and has been
rediscovered recently in the context of the HMF model \cite{bd,hb2,cvb}.

\subsection{The non Markovian kinetic equation}
\label{sec_n}

The above kinetic equations rely on the assumption that the
correlation function relaxes much more rapidly than the distribution
function. The Markovian approximation is expected to be a good
approximation in the limit $N\rightarrow +\infty$ that we consider
since the distribution function changes on a slow timescale of order
$N t_D$ (where $t_D$ is the dynamical time) or even larger. However,
for systems with long-range interactions, there are situations
where the decorrelation time of the fluctuations can be very long so
that the Markovian approximation may not be completely justified.
This concerns in particular the case of self-gravitating systems for
which the temporal correlation of the force decreases like $1/t$ (see
\cite{ct} and Paper II). This is also the case for systems that are
close to the critical point since the exponential relaxation time of
the correlations diverges for $E\rightarrow E_{c}$ or $T\rightarrow
T_{c}$ (see \cite{bouchet,cvb} and Paper II).  Therefore, it can
be of interest to derive non-markovian kinetic equations that may be
relevant to such systems. If we keep the time variation of $f({\bf
v},t)$ in Eq. (\ref{l3}), we obtain after integration
\begin{eqnarray}
\hat{g}({\bf v}_1,{\bf v}_2,{\bf k},t)=\int_0^t d\tau
\frac{i}{m}{\bf k}\hat{u}(k)e^{-i{\bf k}\cdot{\bf w}\tau}\partial
f_1(t-\tau)f_2(t-\tau).\label{n1}
\end{eqnarray}
Inserting this relation in Eq. (\ref{l1}), we obtain a non Markovian
kinetic equation
\begin{eqnarray}
\frac{\partial f_1}{\partial t}=(2\pi)^d m \frac{\partial}{\partial
{v}_1^{\mu}} \int_0^t d\tau \int d{\bf v}_2 d{\bf k} k^{\mu}k^{\nu}
\hat{u}(k)^2 \cos({\bf k}\cdot {\bf w}\tau) \left (\frac{\partial
}{\partial v_1^{\nu}}-\frac{\partial }{\partial v_2^{\nu}}\right
)f({\bf v}_1,t-\tau)f({\bf v}_2,t-\tau). \nonumber\\
\label{n2}
\end{eqnarray}
In particular, for the HMF model, using the notations of Paper I, we
get
\begin{eqnarray}
\frac{\partial f_1}{\partial t}=\frac{k^2}{4\pi}
\frac{\partial}{\partial {v}_1} \int_0^t d\tau \int d{v}_2
\cos\left\lbrack (v_1-v_2)\tau\right\rbrack \left (\frac{\partial
}{\partial v_1}-\frac{\partial }{\partial v_2}\right
)f({v}_1,t-\tau)f({v}_2,t-\tau). \label{n3}
\end{eqnarray}
We note that, when memory terms are taken into account, the collision
term does not vanish. However, if we make the Markovian approximation
$f({v}_1,t-\tau)\simeq f({v}_1,t)$, $f({v}_2,t-\tau)\simeq f({v}_2,t)$
and extend the time integral to infinity \footnote{We could also
consider an approximation where we make the Markovian approximation
but keep the time integral going from $0$ to $t$ (see Appendix
\ref{sec_tempo}).}, we obtain
\begin{eqnarray}
\frac{\partial f_1}{\partial t}=\frac{k^2}{4}
\frac{\partial}{\partial {v}_1}  \int d{v}_2 \delta(v_1-v_2) \left
(\frac{\partial }{\partial v_1}-\frac{\partial }{\partial v_2}\right
)f({v}_1,t)f({v}_2,t)=0.\label{n4}
\end{eqnarray}
When memory terms are neglected we recover the fact that the Landau
collision term vanishes for a spatially homogeneous one-dimensional
system. By working close to the critical point in the HMF model (where
the exponential relaxation time of the correlations diverges), it may
be possible to see non-markovian effects in numerical simulations of
the $N$-body system. They should induce a small evolution of the
homogeneous system on a timescale $Nt_{D}$ as described by
Eq. (\ref{n3}) or, more precisely, by its generalization taking into
account collective effects (see Appendix \ref{sec_tempo}). This should
not lead, however, to statistical equilibrium since Eq. (\ref{n3})
clearly does not tend to the Boltzmann distribution.

\subsection{The kinetic equation for spatially inhomogeneous systems}
\label{sec_g}

Relaxing the assumption that the system is spatially
homogeneous, the equation (\ref{e13}) for the correlation function can be
written
\begin{eqnarray}
\label{g1} {\partial g\over\partial t}+{\cal L} g=-\left \lbrack {\bf {\cal F}}(2\rightarrow 1) {\partial\over\partial {\bf v}_{1}}+{\bf {\cal F}}(1\rightarrow 2) {\partial\over\partial {\bf v}_{2}}\right \rbrack \frac{f}{m}({\bf x}_{1},t)\frac{f}{m}({\bf x}_{2},t),
\end{eqnarray}
where we have denoted the advective term by ${\cal L}$ (Liouvillian
operator). Solving formally this equation with the Green function
\begin{eqnarray}
G(t,t')={\rm exp}\left\lbrace -\int_{t'}^t {\cal
L}(\tau)d\tau,\right\rbrace,\label{g2}
\end{eqnarray}
we obtain
\begin{eqnarray}
g({\bf x}_1,{\bf x}_2,t)=-\int_0^t d\tau G(t,t-\tau) \left \lbrack {\bf {\cal F}}(2\rightarrow 1) {\partial\over\partial {\bf v}_{1}}+{\bf {\cal F}}(1\rightarrow 2) {\partial\over\partial {\bf v}_{2}}\right \rbrack\frac{f}{m}({\bf
x}_1,t-\tau)\frac{f}{m}({\bf x}_2,t-\tau).\nonumber\\
\label{g3}
\end{eqnarray}
The Green function constructed with the smooth field $\langle {\bf
F}\rangle$ means that, in order to evaluate the time integral in Eq.
(\ref{g3}), we must move the coordinates ${\bf r}_i(t-\tau)$ and ${\bf
v}_i(t-\tau)$ of the particles with the mean field flow in phase
space, adopting a Lagrangian point of view. Thus, in evaluating the
time integral, the coordinates ${\bf r}_{i}$ and ${\bf v}_{i}$ placed
after the Greenian must be viewed as ${\bf r}_{i}(t-\tau)$ and ${\bf
v}_{i}(t-\tau)$ where
\begin{eqnarray}
{\bf r}_{i}(t-\tau)={\bf r}_{i}(t)-\int_{0}^{\tau}{\bf v}_{i}(t-s)ds,\quad {\bf v}_{i}(t-\tau)={\bf v}_{i}(t)-\int_{0}^{\tau}\langle {\bf F}\rangle ({\bf r}_{i}(t-s),t-s)ds.
\label{g3c}
\end{eqnarray}
Substituting Eq. (\ref{g2}) in Eq. (\ref{e12}), we get
\begin{eqnarray}
\frac{\partial f_1}{\partial t}+{\bf v}_{1}{\partial f\over\partial {\bf r}_{1}}+\frac{N-1}{N}\langle {\bf F}\rangle_{1} {\partial f\over \partial {\bf v}_{1}}=\frac{\partial}{\partial
{v}_1^{\mu}}\int_0^t d\tau \int d{\bf r}_{2}d{\bf v}_2
{F}^{\mu}(2\rightarrow
1,t)G(t,t-\tau)\nonumber\\
\times  \left \lbrack {{\cal F}}^{\nu}(2\rightarrow 1) {\partial\over\partial { v}_{1}^{\nu}}+{{\cal F}}^{\nu}(1\rightarrow 2) {\partial\over\partial {v}_{2}^{\nu}}\right \rbrack {f}({\bf r}_1,{\bf v}_1,t-\tau)\frac{f}{m}({\bf r}_2,{\bf
v}_2,t-\tau). \label{g4}
\end{eqnarray}
This returns the general kinetic equation obtained by Kandrup
\cite{kandrup1} with the projection operator formalism (note that we
can replace ${F}^{\mu}(2\rightarrow 1,t)$ by ${\cal
F}^{\mu}(2\rightarrow 1,t)$ in the first term of the r.h.s. of the
equation since the fluctuations vanish in average). Equation
(\ref{g4}) slightly differs from the equation obtained in
\cite{kandrup1} by a term $(N-1)/N$ in the l.h.s. This new derivation
of the kinetic equation (\ref{g4}) from a systematic expansion of the
solutions of the BBGKY hierarchy in powers of $1/N$ is valuable
because the present formalism is considerably simpler than the
projection operator formalism and clearly shows which terms have been
neglected in the derivation. It also clearly shows that {\it the
kinetic equation (\ref{g4}) is valid at order $1/N$ so that it
describes the ``collisional'' evolution of the system on a timescale of
order $Nt_D$}.

\subsection{Summary of the different kinetic equations}\label{sec_sum}

Let us briefly summarize the different kinetic equations that
appeared in our analysis. When collective terms are ignored, the
kinetic equation describing the evolution of the system as a whole
at order $1/N$ is
\begin{eqnarray}
\frac{\partial f}{\partial t}+{\bf v}\frac{\partial f}{\partial {\bf
r}}+\frac{N-1}{N}\langle {\bf F}\rangle \frac{\partial f}{\partial
{\bf v}}=\frac{\partial}{\partial v^{\mu}}\int_{0}^{t}d\tau\int
d{\bf r}_{1}d{\bf v}_{1}{F}^{\mu}(1\rightarrow 0)
G(t,t-\tau)\nonumber\\
\times \biggl\lbrace {\cal F}^{\nu}(1\rightarrow
0)\frac{\partial}{\partial v^{\nu}} + {\cal F}^{\nu}(0\rightarrow
1)\frac{\partial}{\partial v_{1}^{\nu}}\biggr\rbrace
{f}({\bf r},{\bf v},t-\tau)\frac{f}{m}({\bf r}_{1},{\bf
v}_{1},t-\tau). \label{sum1}
\end{eqnarray}
If we make a Markov approximation and extend the time integral to infinity, we obtain
\begin{eqnarray}
\frac{\partial f}{\partial t}+{\bf v}\frac{\partial f}{\partial {\bf
r}}+\frac{N-1}{N}\langle {\bf F}\rangle \frac{\partial f}{\partial
{\bf v}}=\frac{\partial}{\partial v^{\mu}}\int_{0}^{+\infty}d\tau\int
d{\bf r}_{1}d{\bf v}_{1}{F}^{\mu}(1\rightarrow 0)
G(t,t-\tau)\nonumber\\
\times \biggl\lbrace {\cal F}^{\nu}(1\rightarrow
0)\frac{\partial}{\partial v^{\nu}} + {\cal F}^{\nu}(0\rightarrow
1)\frac{\partial}{\partial v_{1}^{\nu}}\biggr\rbrace
{f}({\bf r},{\bf v},t)\frac{f}{m}({\bf r}_{1},{\bf
v}_{1},t). \label{sum1b}
\end{eqnarray}
As we have indicated, the Markov approximation is justified for
$N\rightarrow +\infty$ so that $\tau_{corr}\ll t_{relax} \sim
Nt_{D}$. We do not assume, however, that the decorrelation time is
``extremely'' short (i.e, $\tau_{corr}\rightarrow 0$). Therefore, in
the time integral, the distribution functions must be evaluated at
$({\bf r}(t-\tau),{\bf v}(t-\tau))$ and $({\bf r}_{1}(t-\tau),{\bf
v}_{1}(t-\tau))$ where
\begin{eqnarray}
{\bf r}_{i}(t-\tau)={\bf r}_{i}(t)-\int_{0}^{\tau}{\bf v}_{i}(t-s)ds,\quad 
{\bf v}_{i}(t-\tau)={\bf v}_{i}(t)-\int_{0}^{\tau}\langle {\bf F}\rangle ({\bf r}_{i}(t-s),t)ds.
\label{g3cq}
\end{eqnarray}
Comparing Eq. (\ref{g3cq}) with Eq. (\ref{g3c}), we have assumed that
the mean force $\langle{\bf F}\rangle ({\bf r},t)$ does not change
substantially on the timescale $\tau_{corr}$ on which the time
integral has essential contribution. 

For a spatially homogeneous system, using the fact that
Eq. (\ref{g3c}) reduces to ${\bf v}(t-\tau)={\bf v}(t)$ since $\langle
{\bf F}\rangle={\bf 0}$, Eq. (\ref{sum1}) takes the form
\begin{eqnarray}
\frac{\partial f}{\partial t}=\frac{\partial}{\partial
v^{\mu}}\int_{0}^{t}d\tau\int d{\bf r}_{1}d{\bf
v}_{1}{F}^{\mu}(1\rightarrow 0,t) {F}^{\nu}(1\rightarrow 0,t-\tau)\biggl ( \frac{\partial}{\partial
v^{\nu}} -\frac{\partial}{\partial v_{1}^{\nu}}\biggr )
{f}({\bf v},t-\tau)\frac{f}{m}({\bf v}_{1},t-\tau). \label{sum2}
\end{eqnarray}
If we make the integration on ${\bf r}_1$, using the relation (A3) of
Paper II, we obtain the non markovian equation (\ref{n2}). If we make the
Markovian approximation $f({\bf v},t-\tau)\simeq f({\bf v},t)$,
$f({\bf v}_{1},t-\tau)\simeq f({\bf v}_{1},t)$ and extend the time
integration to $+\infty$, we get
\begin{eqnarray}
\frac{\partial f}{\partial t}=\frac{\partial}{\partial
v^{\mu}}\int_{0}^{+\infty}d\tau\int d{\bf r}_{1}d{\bf
v}_{1}{F}^{\mu}(1\rightarrow 0,t) {F}^{\nu}(1\rightarrow
0,t-\tau)\biggl ( \frac{\partial}{\partial v^{\nu}}
-\frac{\partial}{\partial v_{1}^{\nu}}\biggr ) f({\bf
v},t)\frac{f}{m}({\bf v}_{1},t). \label{sum3}
\end{eqnarray}
The integrals on $\tau$ and ${\bf r}_1$ can be performed as in
Appendix A of Paper II and we finally obtain the Landau equation
(\ref{l7}).

\section{Violent collisionless relaxation}
\label{sec_qss}

In this section, we physically discuss the process of violent collisionless
relaxation in relation with the Vlasov equation (\ref{e10}) and point
out several analogies between stellar systems, two-dimensional
turbulence and the HMF model \cite{dauxois}.

\subsection{Quasi Stationary States}
\label{sec_qw}

When the Vlasov equation is coupled to an attractive unshielded
long-range potential of interaction it can develop a process of phase
mixing and violent relaxation leading to the formation of a
quasi-stationary state (QSS). This purely mean field process takes
place on a very short time scale, of the order of a few dynamical
times. This corresponds to the formation of galaxies in astrophysics,
jets and vortices in geophysical and astrophysical flows and clusters
in the HMF model. Lynden-Bell \cite{lb} has proposed to describe these
QSS in terms of statistical mechanics, adapting the usual Boltzmann
procedure so as to take into account the specificities of the Vlasov
equation (in particular the conservation of the infinite class of
Casimirs)
\cite{super}. {\it This approach rests on the assumption that the
collisionless mixing is efficient and that the ergodic hypothesis
which sustains the statistical theory is fulfilled}. There are
situations where the Lynden-Bell prediction works relatively
well. However, there are other situations where the Lynden-Bell
prediction fails. It has been understood since the beginning \cite{lb}
that violent relaxation may be {\it incomplete} in certain cases so
that the Lynden-Bell mixing entropy is {\it not} maximized in the
whole available phase space. Incomplete relaxation \cite{next05} can
lead to more or less severe deviations from the Lynden-Bell
statistics. Physically, the system tries to reach the Lynden-Bell
maximum entropy state during violent relaxation but, in some cases, it
cannot attain it because the variations of the potential, that are the
engine of the evolution, die away before the relaxation process is
complete (there may be other reasons for incomplete relaxation).
Since the Vlasov equation admits an infinite number of stationary
solutions, the coarse-grained distribution $\overline{f}({\bf r},{\bf
v},t)$ can be trapped in one of them $\overline{f}_{QSS}({\bf r},{\bf
v})$ and remain frozen in that quasi stationary state until
collisional effects finally come into play (on longer
timescales). This steady solution is not always the most mixed state
(it can be only partially mixed) so it may differ from Lynden-Bell's
statistical prediction. Thus, for dynamical reasons, the system does
not always explore the whole phase space ergodically.  In general, the
statistical theory of Lynden-Bell gives a relatively good first order
prediction of the QSS without fitting parameter and is able to explain
out-of-equilibrium phase transitions between different types of
structures, depending on the values of the control parameters fixed by
the initial condition. However, there are cases where the prediction
does not work well (it can sometimes be very bad) because of
incomplete relaxation. The difficulty is that we do not know a priori
whether the prediction of Lynden-Bell will work or fail because this
depends on the dynamics and it is difficult to know in advance if the
system will mix well or not.  Therefore, numerical simulations are
necessary to determine how close to the Lynden-Bell distribution the
system happens to be. Let us give some examples of complete and
incomplete violent relaxation in stellar systems, 2D turbulence and
for the HMF model.

\subsection{Stellar systems}
\label{sec_ss}

The concept of violent relaxation was first introduced by Lynden-Bell
\cite{lb} to explain the apparent regularity of elliptical galaxies in
astrophysics. However, for 3D stellar systems the prediction of
Lynden-Bell leads to density profiles whose mass is infinite (the
density decreases as $r^{-2}$ at large distances). In other words,
there is no maximum entropy state at fixed mass and energy in an
unbounded domain. Furthermore, it is known that the distribution
functions (DF) of galaxies do not only depend on the energy
$\epsilon=v^{2}/2+\Phi({\bf r})$ contrary to what is predicted by the
Lynden-Bell statistical theory. This means that other ingredients are
necessary to understand their structure \cite{next05}. However, the approach of
Lynden-Bell is able to explain why elliptical galaxies have an almost
isothermal core. Indeed, it is able to justify a Boltzmannian
distribution $\overline{f}\sim e^{-\beta\epsilon}$ in the core without
recourse to collisions which operate on a much longer timescale $
t_{relax} \sim (N/\ln N)t_{D}$ \cite{chandrastellar}.  By contrast, violent
relaxation is incomplete in the halo. The concept of incomplete
violent relaxation explains why galaxies are more confined than
predicted by statistical mechanics (the density profile of elliptical
galaxies decreases as $r^{-4}$ instead of $r^{-2}$ \cite{bt}). We
note, however, that elliptical galaxies are not stellar polytropes so
their DF cannot be fitted by the Tsallis distribution.

For one dimensional self-gravitating systems, the Lynden-Bell entropy
has a global maximum at fixed mass and energy in an unbounded
domain. Early simulations of the 1D Vlasov-Poisson system starting
from a water-bag initial condition have shown a relatively good
agreement with the Lynden-Bell prediction \cite{hohl}. In other cases,
the Vlasov equation (and the corresponding $N$-body system) can have a very
complicated, non-ergodic, dynamics.  For example, starting from an
annulus in phase space, Mineau {\it et al.} \cite{mineau} have
observed the formation of phase-space holes which block the relaxation
towards the Lynden-Bell distribution. In that case, the system does
not even relax towards a stationary state of the Vlasov equation but
develops everlasting oscillations.

For three dimensional self-gravitating systems confined within a box,
Taruya \& Sakagami \cite{ts} found numerically that the transient
stages of the collisional relaxation of the $N$-stars system can be
fitted by a sequence of polytropic (Tsallis) distributions with a time
dependent $q(t)$ index. Therefore, after the phase of violent
relaxation, the system passes by a succession of quasi-stationary
solutions of the Vlasov equation slowly evolving with time due to
collisions (finite $N$ effects), until the gravothermal catastrophe
associated with the Boltzmann distribution finally takes place.

\subsection{Two-dimensional vortices}
\label{sec_tdv}  

In the context of two-dimensional turbulence, Miller \cite{miller} and
Robert \& Sommeria \cite{rs} have developed a statistical mechanics of
the 2D Euler equation which is similar to the Lynden-Bell theory (see
\cite{csr} for a description of this analogy). This theory works
relatively well to describe vortex merging \cite{rsmepp} or the
nonlinear development of the Kelvin-Helmholtz instability in a shear
layer \cite{staquet}. It can account for the numerous bifurcations
observed between different types of vortices (monopoles, dipoles,
tripoles,...) \cite{jfm1} and is able to reproduce the structure of
geophysical and jovian vortices like Jupiter's great red spot
\cite{bs,turk}.

However, some cases of incomplete relaxation have been reported.  For
example, in the plasma experiment of Huang
\& Driscoll \cite{hd}, the MRS statistical theory
gives a reasonable prediction of the QSS without fit but the agreement
is not perfect \cite{brands}. The observed central density is larger
than predicted by theory and the tail decreases more rapidly than
predicted by theory, i.e. the vortex is more confined. This is related
to the fact that mixing is not very efficient in the core and in the
tail of the distribution (these features can be explained by
developing a kinetic theory of violent relaxation \cite{bbgky}). As
observed by Boghosian \cite{boghosian}, the QSS can be fitted by a
Tsallis distribution where the density drops to zero at a finite
distance.

\subsection{The HMF model}
\label{sec_hmf}

The nature of the quasi stationary states (QSS) observed in the HMF model
has generated an intense (and lively) debate in the community of statistical
physics.

Latora {\it et al.} \cite{latora} performed $N$-body numerical
simulations starting from a water-bag initial condition with
magnetization $M(0)=1$ (unstationary). They observed the formation of
QSS whose lifetime diverges with the system size $N$. The Boltzmann
distribution of statistical equilibrium ${f}_{B}=A e^{-\beta\epsilon}$
is reached on a timescale $t_{relax} \sim N$. In their Fig. 1(a), they
compared the caloric curve $T(U)$ of these QSS (bullets) with the
caloric curve corresponding to the Boltzmann distribution (full
line). They found a range of energies $0.5\lesssim U<U_{c}=3/4$ where
the caloric curve of the QSS disagrees with the caloric curve
corresponding to the Boltzmann statistical equilibrium. We can
interprete their results in another (complementary) manner. First of
all, we note that the caloric curve of the QSS should be compared with
the caloric curve predicted by the Lynden-Bell theory of violent
relaxation since we are dealing with out-of-equilibrium structures
(there is {\it a priori} no reason why the caloric curve of the QSS
resulting from the violent collisionless relaxation should coincide
with the caloric curve of the Boltzmann statistical equilibrium state
resulting from the slow collisional relaxation). The question we now
ask is: can the QSS be described by the Lynden-Bell theory? We note
that, for a water-bag initial condition (two-levels), the distribution
predicted by Lynden-Bell is similar to the Fermi-Dirac statistics
$\overline{f}_{LB}=\eta_{0}/(1+e^{\beta\epsilon+\alpha})$
\cite{super}. Furthermore, for the $M(0)=1$ initial condition, we are
in the dilute (non degenerate) limit of the Lynden-Bell statistical
theory since the initial phase level $\eta_{0}\rightarrow +\infty$
\cite{fermihmf}. Therefore, for  that particular initial condition, 
we remark that the distribution predicted by Lynden-Bell {\it coincides}
with the Boltzmann statistics $\overline{f}_{LB}=A
e^{-\beta\epsilon}$, although it applies to the out-of-equilibrium
QSS.  Because of this coincidence, we can use the caloric curve $T(U)$
reported in Fig. 1(a) of
\cite{latora} to determine the domain of validity of the Lynden-Bell
prediction. This curve shows that the Lynden-Bell prediction works well
for $U>U_{c}=3/4$ (i.e. in the region where the Lynden-Bell maximum
entropy state is spatially homogeneous) and for $U \lesssim 0.5$
(i.e. in the region where the Lynden-Bell maximum entropy state is
strongly spatially inhomogeneous). However, for $0.5\lesssim U<U_{c}$
(i.e. close to the transition energy), the Lynden-Bell prediction
fails. In that case, Latora {\it et al.} \cite{latora} show that the
distribution has the tendency to remain spatially homogeneous
$(M_{QSS}\simeq 0)$ and that the velocity distribution is
non-gaussian.  These results strongly differ from the Lynden-Bell
theory predicting a spatially inhomogeneous state with gaussian
distribution $\overline{f}_{LB}=A e^{-\beta\epsilon}$, the same as the
statistical equilibrium state ${f}_{B}=A
e^{-\beta\epsilon}$. Therefore, close to the critical energy $U_{c}$,
violent relaxation is incomplete and leads to a non-ergodic
behaviour. Since standard statistical mechanics breaks down (standard
statistical mechanics in the present context refers, in our sense, to
the Lynden-Bell theory), Latora {\it et al.}
\cite{latora} propose to describe this regime in terms of Tsallis
generalized thermodynamics. This is an interesting idea to explore
since there are no many other alternatives when the evolution is
non-ergodic (another alternative could be to develop a kinetic theory
of violent relaxation as attempted in \cite{hb4}).

Yamaguchi {\it et al.} \cite{yamaguchi} performed $N$-body numerical
simulations starting from a water-bag initial condition with
magnetization $M(0)=0$. For $U=0.69>U_{c}^{*}=7/12$, this initial
condition is a stable steady state of the Vlasov equation.
Furthermore, we remark that this spatially homogenenous water-bag
initial condition is a minimum of energy $E$ for a given mass $M$ and
phase level value $\eta_{0}$
\cite{fermihmf}. Therefore, this initial condition {\it is} the
Lynden-Bell maximum entropy state. As a result, it does not evolve at
all through the Vlasov equation. Yamaguchi {\it et al.}
\cite{yamaguchi} show that it slowly evolves under the effect of
collisions (finite $N$ effects) by passing through a series of
stationary solutions of the Vlasov equation. This is similar to the
results obtained by Taruya
\& Sakagami \cite{ts} for self-gravitating systems. Yamaguchi {\it et al.} \cite{yamaguchi} show that the Boltzmann distribution is reached on a timescale
$t_{relax} \sim N^{1.7}t_{D}$ and that the velocity distribution of the
transient states is given by the curve reported on their
Fig. 12. Recently, Campa {\it et al.} \cite{campa} obtained similar
results and showed that these transient states can be fitted by a
semi-elliptical distribution. Interestingly, we remark that a
semi-elliptical distribution is a Tsallis distribution $f_{q}(v)=\lbrack
\mu-\beta(q-1)v^{2}/2q\rbrack^{1/(q-1)}$ with an index $q=3$ (if we note
$1-q_{*}$ instead of $q-1$ this corresponds to $q_{*}=-1$)
\footnote{As a result, we can directly apply the stability criterion
of \cite{cvb} to obtain the critical energy above (below) which this
distribution is Vlasov stable (unstable). Note first that the index
$q=3$ corresponds to a polytropic index $n=1$ (see Eq. (144) of
\cite{cvb}) or $\gamma=2$ (see Eq. (145) of
\cite{cvb}). Therefore, according to the stability criterion (156) of \cite{cvb}, 
the critical energy is $\epsilon_{crit}=1/\gamma=1/2$.  Using the
relation $U=\epsilon/4+1/2$ (see Eq. (70) of \cite{fermihmf}) between
the usual energy $U$ used in \cite{latora,yamaguchi,campa} and the
energy $\epsilon$ used in \cite{cvb,fermihmf}, this leads to a critical
energy $U_{crit}=5/8$. This coincides with the result obtained by
Campa {\it et al.} \cite{campa} in a different manner. Note that the
stability criterion (156) of \cite{cvb} which includes the Boltzmann
($\gamma=1$), the Tsallis ($\gamma=1+1/n$ with $n=1/2+1/(q-1)$), the
water-bag ($\gamma=3$) and the semi-elliptical ($\gamma=2$) distributions
 is expressed very simply in terms of the polytropic index
$\gamma$ similar to the one classically used in astrophysics
\cite{bt}.}.  Therefore, we observe that the results 
of Yamaguchi {\it et al.} \cite{yamaguchi} and Campa {\it et al.} 
\cite{campa}, like the results of Taruya \& Sakagami \cite{ts} in
astrophysics, show that {\it Tsallis distributions may be useful to
describe the transient states of a collisional relaxation}. In their
paper, Yamaguchi {\it et al.} \cite{yamaguchi} (see also \cite{bdr})
reject the Tsallis distributions because of the absence of power law
tails in their curves of Fig. 12. However, power law tails are
obtained only for a subclass of Tsallis distributions corresponding to
indices $q<1$ (in our notations \cite{cvb,lang}). For $q>1$, the Tsallis
distributions drop to zero at a finite value of the velocity, so they
have a {\it compact support}. In particular, the distribution obtained
by Yamaguchi {\it et al.}
\cite{yamaguchi} and Campa {\it et al.}
\cite{campa} appears to be well-fitted by a Tsallis
distribution with $q=3>1$ with a tail going to zero abruptly at a
finite velocity $v_{max}$. In view of the lively debate and the
controversy about the applicability of the Tsallis statistics to the HMF
model \cite{yamaguchi,bdr}, it is amusing to realize that the
distribution obtained numerically by Yamaguchi {\it et al.} 
\cite{yamaguchi} is in fact ... a Tsallis distribution! It has a
compact support ($q>1$) instead of power-law tails ($q<1$).

Antoniazzi {\it et al.} \cite{anto} performed $N$-body numerical
simulations starting from a water-bag initial condition with energy
$U=0.69$ and magnetization $M(0)$ between $0$ and $1$. For this value
of energy, the Lynden-Bell theory predicts an out-of-equilibrium phase
transition from a homogeneous state to an inhomogeneous state above a
critical magnetization $M_{crit}=0.897$ discovered in
\cite{fermihmf,anto}. Numerical simulations \cite{anto} show that the
Lynden-Bell prediction works relatively well for
$M(0)<M_{crit}$. This is confirmed by Campa {\it et al.}
\cite{campa} who find in addition that $t_{relax} \sim N^{1.7}t_{D}$ 
as for $M(0)=0$. However, above the critical magnetization, the
results of Latora {\it et al.} \cite{latora} and Campa {\it et al.}
\cite{campa} indicate that the Lynden-Bell theory does not
work since the observed QSS is homogeneous ($M_{QSS}\simeq 0$) with
non-gaussian tails while the Lynden-Bell theory predicts an
inhomogeneous state ($M_{QSS}\neq 0$) with gaussian tails. This
discrepency is particularly clear for the initial condition $M(0)=1$
where the Lynden-Bell distribution coincides with
the Boltzmann distribution $\overline{f}_{LB}\sim e^{-\beta\epsilon}$
(non degenerate limit). Now, the early work of Latora {\it et al.}
\cite{latora} indicates that this gaussian distribution is {\it not}
observed and the recent work of Campa {\it et al.} \cite{campa} (for an isotropic water bag initial condition) shows
that the QSS is well fitted by a semi-elliptical distribution.  As we
have seen, this is a particular Tsallis distribution with index $q=3$
possessing a natural velocity cut-off.  Such distributions, that 
rapidly drop to zero at a finite energy (here velocity) are typical
products of incomplete relaxation. They are explained qualitatively by
the fact that the high energy tail of the distribution in phase space
does not mix well (a similar {\it confinement} is observed in the
plasma experiment of Huang \& Driscoll \cite{hd} discussed in
Sec. \ref{sec_tdv}). This confinement is consistent with a kinetic
theory of incomplete violent relaxation
\cite{next05,bbgky,hb4}. Therefore, as proposed in \cite{fermihmf}, the
out-of-equilibrium phase transition predicted by the Lynden-Bell
theory could be associated with a change of regime in the
dynamics. For $M(0)<M_{crit}$, the system mixes well, the evolution is
ergodic and the violent relaxation is complete leading to the
spatially homogeneous Lynden-Bell distribution. Here, usual
thermodynamics (in the sense of Lynden-Bell) applies.  By contrast,
for $M(0)>M_{crit}$, violent relaxation seems to be incomplete. In
that case, the system does not mix sufficiently well, the evolution is
non ergodic and the observed QSS differs from the Lynden-Bell
prediction. This is associated with the appearance of fractal-like
phase space structures, aging, glassy behaviour, power-law decay of
correlations and anomalous diffusion. Rapisarda \& Pluchino \cite{rp}
have proposed to describe these features in terms of Tsallis
thermodynamics.  On the other hand, in a recent paper, Pluchino {\it
et al.} \cite{prt} have shown explicitly that, in this non-ergodic
regime, time averages and ensemble averages differ. The time averages
can be fitted by $q$-distributions. We propose that the ensemble
averages could also be fitted by a $q$-distribution with an index
$q>1$ leading to a natural velocity cut-off. As we have seen, the
index $q=3$ corresponds to the semi-elliptical distribution observed
by Campa {\it et al.}
\cite{campa}.  Summarizing the above discussion, it seems that the
Lynden-Bell prediction works relatively well far from the transition
line separating homogeneous and inhomogeneous states in the
Lynden-Bell theory (see the phase diagrams reported in
\cite{fermihmf,anto3}) but that it fails close to this transition
line: for fixed $M(0)=1$ this is around $U_{c}=3/4$ (Fig. 1(a) of
\cite{latora} shows a discrepency with the Lynden-Bell theory in that
region) and for fixed $U=0.69$ this is around $M_{crit}=0.897$ (the
simulations of \cite{latora,rp,prt,campa} show a discrepency with the
Lynden-Bell theory for $M(0)\sim 1$). In that case, we have
non-gaussian velocity distributions, phase space structures, glassy
dynamics, aging, anomalous diffusion \footnote{The kinetic theory
developed by Bouchet \& Dauxois \cite{bd} and Chavanis \cite{cvb,hb2}
is valid when the distribution of the bath is spatially {\it
homogeneous}. When the velocity distribution has gaussian tails, like
the Lynden-Bell distributions obtained for $0<M(0)<M_{crit}$
\cite{anto} (the case $M(0)=0$ is special since the
Lynden-Bell distribution coincides with the water-bag distribution
with compact support), the velocity correlation function decays like
$\langle v(0)v(t)\rangle\sim (\ln t)/t$ and the diffusion of angles is
normal (with logarithmic corrections) \cite{bd}. If the velocity
distribution of the bath is water-bag or semi-elliptic, like for
$M(0)=0$
\cite{yamaguchi,campa} or $M(0)=1$
\cite{latora,campa}, standard kinetic theory
\cite{bd,hb2,cvb} predicts that the velocity correlation function has an
exponential decay $\langle v(0)v(t)\rangle\sim e^{-t/\tau}$ leading to
strictly {\it normal} diffusion of angles (this will be checked in a
future contribution). However, if the system exhibits phase space
structures like for $M(0)>M_{crit}$ and relatively small values of $N$
\cite{latora,rp}, the approach of Bouchet \& Dauxois \cite{bd}, {\it
which assumes spatial homogeneity}, is not valid anymore. It is
precisely the presence of these phase space structures that induces
anomalous diffusion as studied in \cite{rp}. Therefore, there should
not be any controversy since these authors \cite{bd} and \cite{rp}
consider {\it different} situations as advocated in
\cite{curious,fermihmf}. }... We must however be very careful because
these striking features, like phase space structures and anomalous
diffusion, could be due to {\it finite size effects}
\cite{moyano,anto} and disappear for $N\rightarrow +\infty$ (note that $N=256000$ in \cite{moyano}, $N=10^5$ in \cite{anto} while $N=2000$ in \cite{rp}). 
In the absence of phase space structures, we suggest that diffusion is
normal because the bath distribution (semi-elliptical
\cite{campa}) has a {\it compact support} (see footnote 5).

Morita \& Kaneko \cite{mk} performed $N$-body numerical simulations
starting from an initial condition with energy $U=0.69$ and
magnetization $M(0)=1$ which is different from the water-bag. In that
case, they find that the system does not relax to a QSS but exhibits
oscillations whose duration diverges with $N$ (they find that the
system relaxes towards the Boltzmann distribution on a timescale
$t_{relax} \sim N$). Therefore, the Lynden-Bell prediction clearly
fails. This long-lasting periodic or quasi periodic collective motion
appears through Hopf bifurcation and is due to the presence of clumps
(high density regions) in phase space. We remark that this behaviour
is relatively similar to the one reported by Mineau {\it et al.}
\cite{mineau} for self-gravitating systems, except that they observe
phase space {\it holes} instead of phase space {\it clumps}.

\section{Slow collisional relaxation}
\label{sec_dis}

In this section, we discuss the process of slow collisional
relaxation in relation with the kinetic equation (\ref{g4}).

\subsection{About the $H$-theorem}

When the system is spatially inhomogeneous, its collisional evolution
can be very complicated and very little is known concerning kinetic
equations of the form (\ref{g4}). For example, it is not
straightforward to prove by a direct calculation that Eq. (\ref{g4})
conserves the energy. However, since Eq. (\ref{g4}) is exact at order
$O(1/N)$, the energy must be conserved. Indeed, the integral
constraints of the Hamiltonian system must be conserved at any order
of the $1/N$ expansion (note that the neglect of collective effects in
Eq. (\ref{g4}) may slightly alter the strict conservation of
energy). On the other hand, we cannot establish the $H$-theorem for an
equation of the form (\ref{g4}). It is only when additional
approximations are implemented (markovian approximation and spatial
homogeneity) that the $H$-theorem is obtained. To be more precise,
let us compute the rate of change of the Boltzmann entropy $S_B=-\int
\frac{f_1}{m}\ln \frac{f_1}{m} d{\bf r}_1 d{\bf v}_1$ with respect to
the general kinetic equation (\ref{g4}). After straightforward
manipulations obtained by interchanging the indices $1$ and $2$, it
can be put in the form
\begin{eqnarray}
\dot S_{B}=\frac{1}{2m^{2}}\int d{\bf x}_{1}d{\bf x}_{2}\frac{1}{f_{1}f_{2}}\int_{0}^{t}d\tau \left\lbrack {\cal F}^{\mu}(2\rightarrow 1)f_2\frac{\partial f_{1}}{\partial v_{1}^{\mu}}+{\cal F}^{\mu}(1\rightarrow 2)f_1\frac{\partial f_{2}}{\partial v_{2}^{\mu}}\right\rbrack_{t} \nonumber\\
\times G(t,t-\tau)\left\lbrack {\cal F}^{\nu}(2\rightarrow 1)f_2\frac{\partial f_{1}}{\partial v_{1}^{\nu}}+{\cal F}^{\nu}(1\rightarrow 2)f_1\frac{\partial f_{2}}{\partial v_{2}^{\nu}}\right\rbrack_{t-\tau}.\label{g6}
\end{eqnarray}
We note that its sign is not necessarily positive. This depends on the
importance of memory effects. In general, the Markovian approximation
is justified for $N\rightarrow +\infty$ because the correlations decay
on a timescale $\tau_{corr}$ that is much smaller than the relaxation
time $t_{relax} \sim Nt_{D}$ on which the distribution changes (as
discussed in Sec. \ref{sec_n}, this approximation is not completely
obvious for self-gravitating systems and for systems that are close to
the critical point). In that case, the entropy increases monotonically
(see Sec. \ref{sec_l} for homogeneous systems). However, even if the
energy is conserved and the entropy increases monotonically, it is not
completely clear whether the general kinetic equation (\ref{g4}) will
relax towards the mean field Maxwell-Boltzmann distribution (I-24) of
statistical equilibrium. It could be trapped in a steady state that is
not the state of maximal entropy because there is no resonance anymore
to drive the relaxation (this is the case for one dimensional
homogeneous systems; see the discussion in Sec. \ref{sec_oned}). It
could also undergo everlasting oscillations without reaching a steady
state.  The kinetic equation (\ref{g4}) may have a rich variety of
behaviors and its complete study is of great complexity.

\subsection{The case of stellar systems}
\label{sec_stellar}

The case of stellar systems is special. These systems are spatially
inhomogeneous but, in order to evaluate the collisional current in
Eq. (\ref{g4}), we can make a {\it local approximation} \cite{bt} and
work as if the system were homogeneous. This is justified by the
divergence of the gravitational force ${\bf F}(2\rightarrow 1)$ when
two particles approach each other so that the fluctuations of the
gravitational force are dominated by the contribution of the nearest
neighbour ${\bf r}_{2}\rightarrow {\bf r}_{1}$ \cite{cvn}. The local
approximation amounts to replacing $f({\bf r}_2,{\bf v}_2,t-\tau)$ by
$f({\bf r}_{1},{\bf v}_2,t-\tau)$ in Eq. (\ref{g4}). This
approximation is justified by the fact that the diffusion coefficient
diverges logarithmically when ${\bf r}_{2}\rightarrow {\bf r}_{1}$
(see below). Using the same argument, we can replace ${\cal
F}^{\mu}(2\rightarrow 1)$ by $F^{\mu}(2\rightarrow 1)$ and ${\cal
F}^{\mu}(1\rightarrow 2)$ by $F^{\mu}(1\rightarrow
2)=-F^{\mu}(2\rightarrow 1)$. We shall also make a markovian
approximation $f({\bf r}_{1},{\bf v}_1,t-\tau)\simeq f({\bf
r}_{1},{\bf v}_1,t)$, $f({\bf r}_{1},{\bf v}_{2},t-\tau)\simeq f({\bf
r}_{1},{\bf v}_{2},t)$ and extend the time integration to
infinity. Then, Eq. (\ref{g4}) becomes
\begin{eqnarray}
\frac{\partial f_1}{\partial t}+{\bf v}_{1}{\partial f\over\partial {\bf r}_{1}}+\frac{N-1}{N}\langle {\bf F}\rangle_{1} {\partial f\over \partial {\bf v}_{1}}=\frac{\partial}{\partial
{v}_1^{\mu}}\int_0^{+\infty} d\tau \int d{\bf r}_{2}d{\bf v}_2
{F}^{\mu}(2\rightarrow
1,t){F}^{\nu}(2\rightarrow
1,t-\tau)\nonumber\\
\times  \left ( {\partial\over\partial { v}_{1}^{\nu}}- {\partial\over\partial {v}_{2}^{\nu}}\right ){f}({\bf r}_1,{\bf v}_1,t)\frac{f}{m}({\bf r}_1,{\bf
v}_2,t).\label{g4bfgr}
\end{eqnarray}
Making a linear trajectory approximation ${\bf
v}_{i}(t-\tau)={\bf v}_{i}(t)$ and ${\bf r}_{i}(t-\tau)={\bf
r}_{i}-{\bf v}_{i}\tau$, we can perform the integrations on ${\bf
r}_{1}$ and $\tau$ like in Appendix A of Paper II. This yields the Vlasov-Landau equation
\begin{eqnarray}
\frac{\partial f_1}{\partial t}+{\bf v}_{1}{\partial f\over\partial {\bf r}_{1}}+\langle {\bf F}\rangle_{1} {\partial f\over \partial {\bf v}_{1}}=\qquad\qquad\qquad\qquad\nonumber\\2\pi m G^{2}\ln\Lambda \frac{\partial}{\partial
{v}_1^{\mu}} \int d{\bf v}_2 \frac{\delta^{\mu\nu}w^{2}-w^{\mu}w^{\nu}}{w^{3}} \left ( {\partial\over\partial { v}_{1}^{\nu}}- {\partial\over\partial {v}_{2}^{\nu}}\right ){f}({\bf r}_1,{\bf v}_1,t){f}({\bf r}_1,{\bf
v}_2,t), \label{g4jfl}
\end{eqnarray}
where $\ln\Lambda=\int_{0}^{+\infty}dk/k$ is the Coulombian factor
\cite{bt}. It must be regularized at small and large scales by
introducing appropriate cut-offs, writing
$\ln\Lambda=\ln(L_{max}/L_{min})$. {\it Note that the divergence at large
scales does not occur in Eq. (\ref{g4})}. It only arises if we assume
that the system is spatially homogeneous and infinite (and if we make
a Markovian approximation and extend the time integral to infinity). In
their stochastic approach, Chandrasekhar \& von Neumann \cite{cvn}
argue that the Coulombian factor must be cut-off at the interparticle
distance because the fluctuations of the gravitational force are
described by the Holtzmark distribution (a particular L\'evy law) that
is dominated by the contribution of the nearest neighbour. However,
Cohen {\it et al.}
\cite{cohen}, considering a Coulombian plasma, argue that the integral 
must be cut-off at the Debye length, which is larger than the
interparticle distance. This is confirmed by the kinetic theory of
Lenard \cite{lenard} and Balescu \cite{balescu} which takes into
account collective effects responsible for Debye shielding. In their
kinetic theory, there is no divergence at large scales and the natural
upper length scale appearing in the Coulombian factor is the Debye
length. In plasmas, a charge is surrounded by a polarization cloud of
opposite charges that diminishes the interaction. For gravitational
systems, there is no shielding so we must stop the integration at $R$,
the system size. Therefore, it is the finite spatial extent of the
system that removes the Coulombian divergence. In a sense, the system
size $R$ (or the Jeans length) plays the role of the Debye length in
plasma physics. On the other hand, the divergence at small scales
comes from the break up of the linear trajectory approximation when
two stars approach each other. This divergence also occurs in
Eq. (\ref{g4}) for the same reason: the unperturbed mean field motion
(\ref{g3c}) becomes incorrect when two stars approach each other. In
fact, to obtain Eq. (\ref{g4}), we have assumed that the correlation
function $g$ is small with respect to $f$. This is true on average,
but it is clear that correlations are important at small scales since
two stars have the tendency to form a binary. {\it Therefore, the
expansion in powers of $1/N$ is not valid at any scale}. One way to
circumvent these difficulties is to use Eq. (\ref{g4}) or
(\ref{g4jfl}) without modification but introduce a cut-off at the
Landau length corresponding to a deflection of $90^{o}$ of the
particles' trajectory \footnote{Note that the divergence at small
scales does not occur in the binary encounter treatment of
Chandrasekhar \cite{chandraB} and Rosenbluth {\it et al.}
\cite{rosenbluth} which takes into account the exact two-body orbit
of the particles instead of making a straight line
approximation.}. Thus, we shall take $L_{min}\sim Gm/v_{typ}^{2}$
where $v_{typ}$ is the typical velocity of a star. Therefore, the
Coulombian factor is estimated by
$\ln\Lambda=\ln(Rv_{typ}^{2}/Gm)$. Now, using a Virial type argument
$v_{typ}^{2}\sim \langle v^{2}\rangle\sim GM/R$, we find that
$\ln\Lambda\sim \ln N$.  The relaxation time $t_{R}$ due to encounters
can be estimated from the Vlasov-Landau equation (\ref{g4jfl}) by
comparing the scaling of the l.h.s. and r.h.s.  This yields
$1/t_{R}\sim mG^{2}\ln\Lambda \rho/v_{typ}^{3}$. The dynamical time is
$t_{D}\sim R/v_{typ}\sim 1/\sqrt{\rho G}$ where $\rho\sim M/R^{3}$ is
the density. Comparing these two expressions, we get the scaling
\begin{eqnarray}
t_{R}\sim \frac{N}{\ln N}t_{D}.
\end{eqnarray}  
A more precise estimate of the relaxation time is given in
\cite{bt,landaud}. The Vlasov-Landau equation conserves the mass, 
the energy (kinetic $+$ potential) and monotonically increases the
Boltzmann entropy. The mean field Maxwell-Boltzmann distribution
(I-24) is the only stationary solution of this equation (cancelling
both the advective term and the collision term
individually). Therefore, the system tends to reach this distribution
on a timescale $(N/\ln N)t_{D}$.  However, there are two reasons why
it cannot attain it: (i) {\it Evaporation:} when coupled to the
gravitational Poisson equation, the mean field Maxwell-Boltzmann
distribution (I-24) yields a density profile with infinite mass so
there is no physical distribution of the form (I-24) in an infinite
domain \cite{paddy,review}.  The system can increase the Boltzmann
entropy indefinitely by evaporating. Therefore, the Vlasov-Landau
equation (\ref{g4jfl}) has no steady state with finite mass and the
density profile tends to spread indefinitely. (ii) {\it Gravothermal
catastrophe:} if the energy of the system is lower than the Antonov
threshold $E_{c}=-0.335GM^{2}/R$ (where $R$ is the system size), it
will undergo core collapse. This is called gravothermal catastrophe
because the system can increase the Boltzmann entropy indefinitely by
contracting and overheating. This process usually dominates over
evaporation and leads to the formation of binary stars
\cite{bt,review}.

\subsection{One dimensional systems}
\label{sec_oned}

One dimensional systems are also special. We have seen in
Sec. \ref{sec_l} that one dimensional systems that are spatially
homogeneous do not evolve at all on a timescale $\sim Nt_{D}$ or
larger because of the absence of resonances. However, if the system is
spatially inhomogeneous, new resonances can appear as described in
\cite{angle} so that an evolution is possible on a timescale
$Nt_{D}$. Then, we can expect that one dimensional inhomogeneous
systems will {\it tend} to approach the Boltzmann distribution on the
timescale $Nt_{D}$.  To be
more precise, let us consider the orbit-averaged-Fokker-Planck
equation derived in
\cite{angle}. Exploiting the timescale separation between the
dynamical time and the relaxation time, we can average Eq. (\ref{g4})
over the orbits, assuming that at any stage of its evolution the
system reaches a mechanical equilibrium on a short dynamical
time. Therefore, the distribution function is a stationary solution of
the Vlasov equation $f\simeq f(\epsilon,t)$ [where
$\epsilon=v^{2}/2+\Phi$ is the individual energy] slowly evolving in
time under the effect of ``collisions'' ($=$ correlations due to
finite $N$ effects). Introducing angle-action variables, we get an
equation of the form \cite{angle}:
\begin{equation}
\label{q16}
\frac{\partial f}{\partial t}=\frac{1}{2}\frac{\partial}{\partial J}\sum_{m,m'}\int {m A_{mm'}(J,J')^{2}}\delta(m\Omega(J)-m'\Omega(J'))\left\lbrace f(J')m\frac{\partial f}{\partial J}-f(J)m'\frac{\partial f}{\partial J'}\right\rbrace dJ'.
\end{equation} 
The important point to notice is that the evolution of the system is
due to a condition of resonance between the pulsations $\Omega(J)$ of
the particles' orbits (this property probably extends to $d$
dimensions but is technically more complicated to show). Only
particles whose pulsations satisfy $m\Omega(J)=m'\Omega(J')$ with
$(m,J)\neq (m',J')$ participate to the diffusion current. This is
similar to the collisional relaxation of two dimensional point
vortices
\cite{chavlemou,bbgky}. It can be shown that Eq. (\ref{q16})
conserves mass and energy and monotonically increases entropy so that
the system {\it tends} to approach the Boltzmann distribution of
statistical equilibrium on a timescale $\sim Nt_{D}$
\cite{angle}. However, it may happen that there is not enough
resonances so that the system can be trapped in a quasi stationary
state different from the Boltzmann distribution. This happens when the
condition of resonance cannot be satisfied so that $m\Omega(J)\neq
m'\Omega(J')$ for all $(m,J)\neq (m',J')$. In that case, the system is
in a steady state of Eq. (\ref{q16}) which is not the Boltzmann
distribution.  This is what happens to point vortices in 2D
hydrodynamics when the profile of angular velocity becomes monotonic
\cite{chavlemou}.  In that case, the relaxation stops and the system
will relax on a timescale larger than $Nt_{D}$. We may wonder whether
the same situation can happen to systems described by a kinetic
equation of the form (\ref{g4}).

\section{Conclusions and perspectives}
\label{sec_conc}

In this paper, we have developed a kinetic theory for Hamiltonian
systems with weak long-range interactions. A specificity of these
systems is that they can be spatially inhomogeneous, which considerably
complicates the kinetic theory.  We have shown that the developement
of correlations between particles creates a current in the r.h.s. of
Eq. (\ref{e6}) that is the counterpart of the collision term in the
Boltzmann equation for neutral gases. Therefore, for Hamiltonian
systems with weak long-range interactions, the evolution beyond the Vlasov
regime is driven by ``correlations'' due to finite $N$ effects. We
have obtained a kinetic equation (\ref{g4}) valid at order $O(1/N)$
that describes the evolution of the system on a timescale $\sim
Nt_{D}$. For homogeneous systems, this equation reduces to the Landau
equation. For $d>1$, the Landau equation relaxes towards the Boltzmann
distribution. Therefore, for $d>1$, the relaxation time scales like
$t_{relax} \sim Nt_{D}$.  This scaling has been predicted and observed
in a spatially homogeneous two-dimensional Coulombian plasma
\cite{hb2,benedetti,landaud}. This scaling probably remains true for
spatially inhomogeneous systems in $d>1$ with the exception of
self-gravitating systems that relax towards the mean-field Boltzmann
distribution on a timescale $t_{relax} \sim (N/\ln N)t_{D}$, unless they
experience evaporation or gravothermal catastrophe. For one
dimensional systems, like the HMF model, the situation is more
complicated. For Vlasov-stable homogeneous systems, the kinetic
equation (\ref{g4}) reduces to $\partial f/\partial t=0$. Therefore,
there is no evolution on a timescale of the order $\sim Nt_{D}$. We
conclude that the relaxation time is larger than $Nt_{D}$. We could
imagine that the evolution is due to three-body,
four-body,... correlations leading to a relaxation time of the order
of $N^{2}t_{D}$, $N^{3}t_{D}$,... However, Campa {\it et al.}
\cite{campa}, considering initial conditions with supercritical energy
$U>U_{c}=3/4$ for which the system is always spatially homogeneous,
found that the relaxation time is extremely long scaling like
$t_{relax} \sim e^{N}$. This suggests that the expansion of the BBGKY
hierarchy in powers of $1/N$ may not be convergent in the homogeneous
case and that another approach should be developed in that case. On
the other hand, Morita \& Kaneko \cite{mk} considering an initial
condition with $U<U_{c}$ and $M(0)=1$, found a relaxation time of the
order $t_{relax} \sim Nt_{D}$. In their simulations, the system is
always spatially inhomogenous (the magnetization in the oscillatory
regime is non-zero). As explained in Sec. \ref{sec_oned}, spatial
inhomogeneities can create new resonances that drive the relaxation
towards the Boltzmann equilibrium (BE) on a timescale $t_{relax} \sim
Nt_{D}$ (predicted by the kinetic theory) that is much shorter than
when the system remains spatially homogeneous. This could be an
explanation (but not the only one) for the observed timescale in \cite{mk}.
Yamaguchi {\it et al.}
\cite{yamaguchi}, considering a water-bag initial condition with $U<U_{c}$
and $M(0)=0$, found a relaxation time $t_{relax} \sim N^{\delta} t_{D}$
with $\delta=1.7$.  In their simulations, the system is spatially
homogeneous but it progressively becomes Vlasov unstable and undergoes
a {\it dynamical phase transition} from the homogeneous QSS to the
inhomogeneous BE. This instability considerably accelerates
the relaxation towards the Boltzmann equilibrium with respect to the
case $U>U_{c}$ \cite{campa} where the homogeneous distribution remains Vlasov
stable until the end (see below). The same phase transition happens in
the simulations of Latora {\it et al.}
\cite{latora} who considered a water-bag initial condition with
$U<U_{c}$ and $M(0)=1$. Their system is roughly spatially homogeneous
($M_{QSS}\simeq 0$) but it also presents some phase space structures
which may explain why they find a relaxation time of the order
$t_{relax} \sim Nt_{D}$ shorter than $t_{relax} \sim N^{1.7} t_{D}$
(we have seen that spatial inhomogeneities can accelerate the
relaxation by creating new resonances).  More generally, for water-bag initial
conditions, it would be interesting to determine how the exponent
$\delta$ depends on the initial magnetization $M(0)$ and energy
$U$. Considering the phase diagram in $(M(0),U)$ plane reported in
\cite{anto3}, we suggest that $t_{relax} \sim e^N$ above the critical line $U_{c}=3/4$ (no dynamical phase transition) and $t_{relax} \sim N^{\delta} t_{D}$  below the critical line $U_{c}=3/4$ (dynamical phase transition). 
For given initial magnetization $M(0)$, we expect that $\delta$
diverges as we approach the critical energy $U_{c}=3/4$ above which
$t_{relax} \sim e^{N}$. Below the critical line $U_{c}=3/4$ and above
the transition line $U_{crit}(M(0))$ (the curve in Fig. 1 of
\cite{anto3}), observations \cite{yamaguchi} 
show that $\delta\sim 1.7$. Below the transition line
$U_{crit}(M(0))$, the QSS is either spatially inhomogeneous (according
to Lynden-Bell's prediction if relaxation is complete) or spatially
homogeneous with phase space structures in case of incomplete
relaxation \cite{latora,rp}.  In that case, observations \cite{latora}
show that $\delta\sim 1$, which is consistent with the kinetic theory
for inhomogeneous systems. Thus, for a given energy $U<U_{c}$, we
expect that $\delta$ passes from $\delta>1.7$ to $\delta=1$ when
$M(0)$ overcomes the critical magnetization $M_{crit}(U)$
\cite{fermihmf,anto3}. These ideas, that are
consistent with the partial numerical information that we have at
present, demand to be developed in more detail. In fact, this is a
first attempt to connect the relaxation time to the phase diagram
$(U,M(0))$ and the situation may be more complicated than that. In
particular, the relaxation time seems to strongly depend on the
detailed structure of the inital condition. For example, using an {\it
isotropic} water bag initial condition with $M(0)=1$, Campa {\it et
al.} \cite{campa} find a relaxation time $t_{relax}\sim N^{1.7}t_{D}$
instead of the scaling $t_{relax}\sim N t_{D}$ reported by Latora {\it
et al.} \cite{latora}. This may be related to the lack of phase space
structures in the simulations of \cite{campa}.

Finally, we conclude by proposing the following scenario similar to
the one proposed by Taruya \& Sakagami \cite{ts} for self-gravitating
systems. Analyzing the numerical results of
\cite{latora,yamaguchi,campa}, we argue that, for many initial
conditions, the transient states of the collisional relaxation of the
HMF model can be described by spatially homogeneous Tsallis
distributions (polytropes) with a time varying index $q(t)\ge 1$
(i.e. $n(t)\ge 1/2$, $1\le
\gamma(t)\le 3$) corresponding to a compact support. More precisely, we 
parametrize these transient
states by a distribution function of the form $f(v,t)=f(0,t)\lbrack
1-v^{2}/v_{max}(t)^{2}\rbrack^{n(t)-1/2}$. It is easy to show that
$n(t)$ and $v_{max}(t)$ are related to each other by $n(t)=2\pi
v_{max}(t)^{2}/(kN\epsilon)-1$ where $\epsilon=4(U-1/2)$ is the
conserved energy and $2\pi/(kN)=1$ in usual notations. This simple
relation follows from Eqs. (28) and (29) of \cite{lang} for
homogeneous systems where $\Phi=0$, $\rho=N/2\pi$ and $p=E/\pi$. It
allows to determine $n(t)$ by simply measuring $v_{max}(t)$.  These
Tsallis distributions are quasi-stationary solutions of the Vlasov
equation slowly evolving with time under the effect of collisions
(finite $N$ effects). Initially, $n(t)$ is close to $n=1$
(i.e. $\gamma=2$, $q=3$) corresponding to the semi-elliptical
distribution observed by Campa {\it et al.}
\cite{campa} in many circumstances. For $U=0.69$ 
(i.e. $\epsilon=0.76$) and $n=1$ we get $v_{max}\simeq
1.23$. Progressively, $n(t)$ increases so as to attain the value
$n\rightarrow +\infty$ (i.e. $\gamma=1$, $q=1$) corresponding to the
Boltzmann equilibrium state for $t\rightarrow +\infty$. For
$\epsilon<1$ (i.e. $U<U_{c}=3/4$), there exists a time $t_{*}$ at
which $n(t)=n_{crit}=\epsilon/(1-\epsilon)$ (corresponding to
$\gamma(t)=\gamma_{crit}=1/\epsilon$,
$q(t)=q_{crit}=(\epsilon+1)/(3\epsilon-1)$) so that the Tsallis distribution
becomes Vlasov unstable (see the criterion (156) of \cite{cvb}) and
the system rapidly relaxes towards the inhomogeneous Boltzmann
distribution \footnote{Note that Taruya \& Sakagami
\cite{ts} interprete this transition  as a generalized
thermodynamical instability (in Tsallis sense) while {\it we interprete it
as a dynamical instability with respect to the Vlasov equation
\cite{lang}}.}. This accounts for the sudden dynamical phase transition observed in \cite{yamaguchi} from the
homogeneous QSS ($M_{QSS}\simeq 0$) to the inhomogeneous BE
($M=M_{eq}\neq 0$). For $U=0.69$, the transition corresponds to
$n_{crit}\simeq 3.166$ leading to $v_{max}\simeq 1.78$ in qualitative
agreement with \cite{campa}. In the supercritical case $\epsilon>1$
(i.e. $U>U_{c}=3/4$), the Tsallis distributions with $q(t)\ge 1$ are
{\it always} Vlasov stable (since
$q_{crit}=(\epsilon+1)/(3\epsilon-1)=(4U-1)/(12U-7)<1$ or,
alternatively, $\epsilon_{crit}=(q+1)/(3q-1)<1$ or
$U_{crit}=(7q-1)/\lbrack 4(3q-1)\rbrack<U_{c}=3/4$) {\it which
explains the long lifetime behaviour observed by Campa {\it et al.}
\cite{campa}}.  This scenario suggests that Tsallis distributions can be {\it attractors} (or at least provide a good fit) for the transient
states of the collisional relaxation, for a large class of initial
conditions ($U=0.69<U_{c}$ with $M(0)=0$ \cite{yamaguchi,campa};
$U=0.69<U_{c}$ with $M(0)=1$ \cite{latora,campa}; and $U>U_{c}$
\cite{campa}).  Note, however, that the previous scenario is not valid
for all initial conditions so these attractors are {\it not universal}. For
example, in the numerical simulations of Antoniazzi {\it et al.} 
\cite{anto}, the system relaxes towards a Lynden-Bell distribution with gaussian tails for $0<M<M_{crit}=0.897$ and in the numerical simulations of Morita \&
Kaneko \cite{mk}, the system develops everlasting
oscillations. However, this picture now suggests to look in detail into
the chaotic dynamics of the system in order to determine the basin of
attraction of the Tsallis distributions \cite{latora,yamaguchi,campa},
the Lynden-Bell distributions \cite{anto} and the oscillatory states
\cite{mk}. We hope to develop these issues, and check the above
scenario, in future communications.

\appendix

\section{Some other kinetic equations}
\label{sec_tempo}

If we make the Markovian approximation $f({\bf v}_1,t-\tau)\simeq
f({\bf v}_1,t)$ and $f({\bf v}_2,t-\tau)\simeq
f({\bf v}_2,t)$ in Eq. (\ref{n2}) but do not extend the time integral to infinity, we obtain
\begin{eqnarray}
\frac{\partial f_1}{\partial t}=\frac{\partial}{\partial
{v}_1^{\mu}} \int d{\bf v}_2  K^{\mu\nu}({\bf w},t) \left (\frac{\partial
}{\partial v_1^{\nu}}-\frac{\partial }{\partial v_2^{\nu}}\right
)f({\bf v}_1,t)f({\bf v}_2,t), \nonumber\\
\label{tempo1}
\end{eqnarray}
with
\begin{eqnarray}
K^{\mu\nu}({\bf w},t)=(2\pi)^d m \int_0^t d\tau \int d{\bf k} k^{\mu}k^{\nu}
\hat{u}(k)^2 \cos({\bf k}\cdot {\bf w}\tau).
\label{tempo2}
\end{eqnarray}
In $d=3$, the components of this tensor can be calculated by introducing a
spherical system of coordinates with the $z$-axis in the direction of
${\bf w}$. We find that
\begin{eqnarray}
K^{\mu\nu}({\bf w},t)=A(w,t)\frac{w^{2}\delta^{\mu\nu}-w^{\mu}w^{\nu}}{w^{2}}+
B(w,t)\frac{w^{\mu}w^{\nu}}{w^{2}},
\label{tempo3}
\end{eqnarray}
with
\begin{eqnarray}
A(w,t)=\frac{8\pi^4 m}{w}\int_{0}^{+\infty}k^{3}dk \hat{u}(k)^{2}\int_{0}^{kwt}\left (\frac{4\sin\tau}{\tau^{3}}-\frac{4\cos\tau}{\tau^{2}}\right )d\tau,
\label{tempo4}
\end{eqnarray}
\begin{eqnarray}
B(w,t)=\frac{16\pi^4 m}{w}\int_{0}^{+\infty}k^{3}dk \hat{u}(k)^{2}\int_{0}^{kwt}\left (\frac{2\sin\tau}{\tau}+\frac{4\cos\tau}{\tau^{2}}-\frac{4\sin\tau}{\tau^{3}}\right )d\tau.
\label{tempo5}
\end{eqnarray}
For $t\rightarrow +\infty$, the functions $A$ and $B$ reduce to 
\begin{eqnarray}
A(w)=\frac{8\pi^5 m}{w}\int_{0}^{+\infty}k^{3} \hat{u}(k)^{2}dk, \qquad B(w)=0,
\label{tempo6}
\end{eqnarray}
and we recover the results (II-42) and (II-43) of Paper II. In $d=2$,
the components of the tensor (\ref{tempo2}) can be calculated by
introducing a polar system of coordinates with the $x$-axis in the
direction of ${\bf w}$. This leads to Eq. (\ref{tempo3}) with now
\begin{eqnarray}
A(w,t)=\frac{8\pi^{3}m}{w}\int_{0}^{+\infty}k^{2}dk \hat{u}(k)^{2}\int_{0}^{kwt}\frac{J_{1}(\tau)}{\tau}d\tau,
\label{tempo7}
\end{eqnarray}
\begin{eqnarray}
B(w,t)=\frac{8\pi^{3}m}{w}\int_{0}^{+\infty}k^{2}dk \hat{u}(k)^{2}\int_{0}^{kwt}\left \lbrack \frac{J_{1}(\tau)}{\tau}-J_{2}(\tau)\right\rbrack d\tau.
\label{tempo8}
\end{eqnarray}
For $t\rightarrow +\infty$, the functions $A$ and $B$ reduce to 
\begin{eqnarray}
A(w)=\frac{8\pi^3 m}{w}\int_{0}^{+\infty}k^{2} \hat{u}(k)^{2}dk, \qquad B(w)=0,
\label{tempo9}
\end{eqnarray}
and we recover the results (II-42) and (II-43) of Paper II. Finally,
in $d=1$, we obtain
\begin{eqnarray}
K(w,t)=4\pi m\int_{0}^{+\infty} k^{2} \hat{u}(k)^{2}\frac{\sin(kwt)}{kw}dk.
\label{tempo10}
\end{eqnarray}
For $t\rightarrow +\infty$, we find that
\begin{eqnarray}
K(w)=4\pi^{2} m \delta(w)\int_{0}^{+\infty} k \hat{u}(k)^{2}dk,
\label{tempo11}
\end{eqnarray}
which returns Eq. (\ref{l8}).
Finally, we recall that Eq. (\ref{n2}) ignores collective effects. As
a simple generalization, we could replace in Eq. (\ref{n2}) the
potential $\hat{u}(k)$ by the ``screened'' potential
$\hat{u}(k)/|\epsilon({\bf k},{\bf k}\cdot {\bf v}_{2})|$ including
the dielectric function (for the HMF model, this amounts to dividing
the integrand of Eq. (\ref{n3}) by
$|\epsilon(1,v_{2})|^{2}$). This leads to 
\begin{eqnarray}
\frac{\partial f_1}{\partial t}=(2\pi)^d m \frac{\partial}{\partial
{v}_1^{\mu}} \int_0^t d\tau \int d{\bf v}_2 d{\bf k} k^{\mu}k^{\nu}
\frac{\hat{u}(k)^2}{|\epsilon({\bf k},{\bf k}\cdot {\bf v}_{2})|^{2}} \cos({\bf k}\cdot {\bf w}\tau) \nonumber\\
\times\left (\frac{\partial
}{\partial v_1^{\nu}}-\frac{\partial }{\partial v_2^{\nu}}\right
)f({\bf v}_1,t-\tau)f({\bf v}_2,t-\tau). 
\label{tempo12}
\end{eqnarray}
Strictly speaking, this procedure is not rigorously justified for non
Markovian systems since the dielectric function is obtained by
assuming precisely that the distribution function does not change on
the timescale of interest. Yet, this generalization could be performed
heuristically in order to obtain a non Markovian kinetic equation
taking into account some collective effects. If we make the Markovian
approximation and extend the time integral to infinity, Eq. (\ref{tempo12})
returns the Lenard-Balescu equation.


\end{document}